\documentclass[aps,prl,twocolumn,superscriptaddress,showpacs,floatfix]{revtex4}
\pdfoutput=1

\usepackage{graphicx}
\usepackage{dcolumn}
\usepackage{bm}
\usepackage{stmaryrd}
\usepackage{latexsym}
\usepackage{amssymb}
\usepackage{amsfonts}
\usepackage{amsmath}

\begin{document}
\title{Time-dependent current-density-functional theory of spin-charge separation and spin drag in one-dimensional ultracold Fermi gases}
\date{\today}

\author{Gao Xianlong}
\affiliation{Department of Physics, Zhejiang Normal University, Jinhua, Zhejiang Province, 321004, China}
\author{Marco Polini}
\email{m.polini@sns.it}
\affiliation{NEST-CNR-INFM and Scuola Normale Superiore, I-56126 Pisa, Italy}
\author{Diego Rainis}
\affiliation{NEST-CNR-INFM and Scuola Normale Superiore, I-56126 Pisa, Italy}
\author{M.P. Tosi}
\affiliation{NEST-CNR-INFM and Scuola Normale Superiore, I-56126 Pisa, Italy}
\author{G. Vignale}
\affiliation{Department of Physics and Astronomy, University of
Missouri, Columbia, Missouri 65211, USA}

\begin{abstract}
Motivated by the large interest in the non-equilibrium dynamics of low-dimensional quantum many-body systems, we present a fully-microscopic theoretical and numerical study of the ``charge" and ``spin" dynamics in a one-dimensional ultracold Fermi gas following a quench. Our approach, which is based on time-dependent current-density-functional theory, is applicable well  beyond the linear-response regime and 
produces both spin-charge separation and spin-drag-induced broadening of the spin packets.
\end{abstract}
\maketitle

\noindent {\it Introduction}---The dynamics of low-dimensional quantum many-body systems is a topic of great theoretical and experimental interest.  
With the recently acquired capability to store and control ultracold atomic gases~\cite{rmp_cold_atoms}, it has become possible to study in the laboratory some ideal systems which were once accessible only to theoretical investigation.  In particular, it is now possible to confine atoms along a waveguide, thus realizing for the first time the elusive one-dimensional Fermi and Bose gases~\cite{1D_cold_atoms}.

One-dimensional (1D) Fermi systems are theoretically interesting because they present us with a breakdown of the Landau Fermi liquid paradigm~\cite{giamarchi_book}. The low-energy excitations of such systems are collective, while those of a conventional Fermi liquid are single-particle like; furthermore there is complete separation between charge and spin excitations, which in an ordinary Fermi liquid are tied together in the ``quasiparticle".  

Fully-microscopic calculations of collective spin and charge dynamics in
1D Fermi systems have recently been performed by Kollath {\it et al.}~\cite{kollath_prl_2005} by means of the time-dependent density-matrix renormalization-group method. This powerful numerical method seems however to be limited to inhomogeneous lattice systems at zero temperature.  
In Ref.~\onlinecite{polini_2007} we have pointed out a novel aspect of the 1D spin-charge separation phenomenon, which appears only at finite temperature.  Namely, we have shown that the propagation of a spin-density packet becomes intrinsically diffusive due to the {spin drag effect}~\cite{scd_giovanni,scd_flensberg,weber_nature_2005} -- the transfer of momentum between fermions of opposite spin orientations.  We have developed a {\it macroscopic} approach, formulated entirely in terms of collective variables, for calculating the propagation of spin-density or just density packets, and we believe that the difference between the two cases can be experimentally detected.

However, some difficulties persist, which limit our ability to calculate theoretically the dynamics of spin and density packets: (i) single-particle thermal excitations, which may affect the shape of a spin-density packet in ways that are difficult to distinguish from the effect of the spin drag, are beyond the reach of the macroscopic treatment of  Ref.~\onlinecite{polini_2007}; 
(ii) the macroscopic theory, being restricted to long wavelengths and low frequencies, misses  the interaction of the wavepacket with microscopic Friedel-like oscillations, which are induced, for example, by the sharp ends of the waveguide; and, finally, (iii) the macroscopic theory is restricted to the linear-response regime (LRR), whereas any experiment that is likely to be performed with ultracold atomic gases 
would start with a strong local disturbance~\cite{kollath_prl_2005}. 

It might appear that these difficulties are formidable enough to prevent any further progress, but it is not so.  It turns out that a theoretical tool already exists, which can take care of all three problems in a relatively simple manner.  It is called  {\it ``time-dependent spin-current density functional theory"} (TD-SCDFT)~\cite{qian_prl_2003,TDFTBook06,Giuliani_and_Vignale} and it is based on the idea of mapping the time-dependent many-body problem into an effective single-particle problem, with an effective potential that is designed to produce the correct evolution of the spin density and the spin-current density.  

Single-particle aspects are naturally included in TD-SCDFT 
because the spin-resolved densities are represented as a sum of contributions from 
one-particle wavefunctions, 
which are fully microscopic objects that are populated according to a Fermi-Dirac distribution function 
(so thermal effects and Friedel oscillations are automatically taken into account).  Collective effects are included through an exchange-correlation (xc) field, which is expressed in terms of collective variables -- the spin density and the spin-current density -- and is derived non-empirically from a suitable homogeneous reference system. Finally, the theory is not restricted to the LRR.

In this paper we present the first application of  TD-SCDFT to the study of the collective dynamics of wavepackets in 1D Fermi gases, and we compare the results with those of 
Ref.~\onlinecite{polini_2007}.   We show that the simplest xc potential, constructed from the adiabatic local-spin-density approximation (ALSDA), is already sufficient to produce  spin-charge separation.  But we also observe that the density packet gets progressively ``contaminated" with microscopic Friedel-like oscillations coming from the ends of the waveguide, which were absent in the macroscopic calculation.  Going beyond the ALSDA, we show that the spin-drag effect is nicely captured and continues to be the dominant cause of diffusion.
In short, we demonstrate the feasibility of a novel and fully microscopic computational approach,  which leads to detailed predictions to be compared with future experiments.  

\noindent {\it The model}---We consider a two-component Fermi gas ({\it e.g.} 
a mixture of two hyperfine states of $^6$Li~\cite{spin_pol_fermions}) 
with $N$ atoms confined inside a tight waveguide 
along the ${\hat x}$ direction. The two species of 
atoms are assumed to have the same mass $m$ and different  ``spin"  (hyperfine state label) $\sigma=\uparrow$ or $\downarrow$. The fermions interact  {\it via} a zero-range $s$-wave repulsive potential $v(x)=g_{\rm 1D}\delta(x)$~\cite{olshanii_1998}, whose strength can be tuned {\it e.g.} by using a magnetic field-induced Feshbach resonance between the two different spin states~\cite{1D_cold_atoms}.
The Fermi gas is confined in the waveguide by a static trap $V(x)$.

Similarly to what done in Refs.~\onlinecite{kollath_prl_2005} and~\onlinecite{scsep_1D_cold_gases}, 
we now imagine that a local external potential, which couples only to atoms of one spin species (hereby referred to as ``{\it the accumulator}"), acts on the Fermi gas for times $t \leq 0$ creating an accumulation of, say, up-spin atoms close to the trap center. The state of the system for times $t \leq 0$ is denoted by $|\Psi_0\rangle$, which 
is the ground state of the Hamiltonian 
${\hat {\cal H}}_0={\hat {\cal H}}_{\rm Y}+{\hat V}+{\hat V}_{\rm acc}$. Here 
${\hat {\cal H}}_{\rm Y}=\sum_{i}{\hat p}^2_i/(2m)+g_{\rm 1D}\sum_{i<j}\delta({\hat x}_i-{\hat x}_j)$ is the Yang Hamiltonian~\cite{yang_1967}, 
${\hat V}=\sum_i V({\hat x}_i)$ describes the trapping potential,
and, finally, ${\hat V}_{\rm acc}=\sum_{i, \sigma} W_\sigma({\hat x}_i)$ with 
$W_\sigma(x)=A_\sigma \exp{[-x^2/(2w^2)]}$, is the potential that describes the accumulator~\cite{accum} 
($A_\uparrow<0$ and $A_\downarrow=0$). 
We do not make any assumption on the strength $A_\uparrow$ of the accumulator ({\it i.e.} we do not assume to be in the LRR). 
 
At time $t=0^+$ the accumulator is {\it suddenly} turned off and the ``charge", $n(x,t)=n_\uparrow(x,t)+n_\downarrow(x,t)$, and ``spin",  $s(x,t)=n_\uparrow(x,t)-n_\downarrow(x,t)$, densities are allowed to propagate along the waveguide in the presence of $V(x)$. Here $n_\sigma(x,t)= \langle \Psi(t)| {\hat \psi}^\dagger_\sigma(x){\hat \psi}_\sigma(x) |\Psi(t) \rangle$ are the spin-resolved densities, 
with $|\Psi(t)\rangle$ the state of the system at time $t$~\cite{time_independence} and ${\hat \psi}^\dagger_\sigma(x)$ a field-operator that creates 
a fermion with spin $\sigma$ at position $x$. 

\noindent {\it Dynamics from TD-SCDFT}---According to TD-SCDFT~\cite{qian_prl_2003,TDFTBook06,Giuliani_and_Vignale}, 
the spin-resolved densities $n_\sigma(x,t)$ and the associated current densities $j_\sigma(x,t)$ at times $t>0$ 
can be found by solving the time-dependent Kohn-Sham (KS) equations
\begin{equation}\label{eq:kss}
i\hbar\partial_t \psi_{\alpha,\sigma}(x,t)=\left[-\frac{\hbar^2}{2m}\partial^2_x+V^{(\sigma)}_{\rm KS}[n_\sigma, j_\sigma](x,t)\right]
\psi_{\alpha, \sigma}(x,t)\,,
\end{equation}
where $V^{(\sigma)}_{\rm KS}[n_\sigma, j_\sigma](x,t)=V(x)+V^{(\sigma)}_{\rm H}[n_\sigma](x,t)+V^{(\sigma)}_{\rm xc}[n_\sigma,  j_\sigma](x,t)$ is the KS potential, which includes the trapping potential $V$,  the Hartree mean-field potential  [$V^{(\sigma)}_{\rm H}[n_\sigma](x,t)=g_{\rm 1D} n_{\bar \sigma}(x,t)$,  where ${\bar \sigma}=-\sigma$], and the xc potential -- the latter a functional of the spin and current densities.  Although, in general, the xc effects in TD-SCDFT are represented by a {\it vector potential}~\cite{TDFTBook06}, it turns out that in 1D a vector potential can be transformed into a scalar potential by an  appropriate gauge transformation: here we have already taken advantage of this possibility. 
 The  densities are self-consistently determined via the usual relation 
\begin{equation}\label{eq:closure}
n_\sigma(x,t)=\sum_{\alpha} \frac{\left|\psi_{\alpha, \sigma}(x,t)\right|^2}
{\exp{[(\varepsilon_{\alpha, \sigma}-\mu)/(k_{\rm B} T)]}+1}~, 
\end{equation}
where $\varepsilon_{\alpha,\sigma}$ are the static KS energies of the initial state   with chemical potential $\mu$.
These energies are found by solving a static KS self-consistent problem corresponding to ${\hat {\cal H}}_0$.  The current densities are related to the densities by the continuity equations $\partial_x j_\sigma(x,t) = -\partial_t n_\sigma(x,t)$.
Note that due to the time-dependence of $n_\sigma$ and $j_\sigma$, the KS Hamiltonian is time-{\it dependent}.

In order to proceed we need a sensible approximation for the xc potential.  In order to grasp the physically relevant facts that occur after the quench described above, we make use of the following  approximate  expression: 
$V^{(\sigma)}_{\rm xc}[n_\sigma, j_\sigma](x,t)\simeq V^{(\sigma)}_{\rm ALSDA}[n_\sigma](x,t)+V^{(\sigma)}_{\rm dyn}[n_\sigma, j_\sigma](x,t)$. In this equation $V^{(\sigma)}_{\rm ALSDA}[n_\sigma](x,t)$ represents the ALSDA contribution, which depends only on the densities~\cite{zeroT}:
\begin{equation}\label{eq:alsda}
V^{(\sigma)}_{\rm ALSDA}[n_\sigma](x,t) = \left.\frac{\partial [n \varepsilon^{\rm hom}_{\rm xc}(n_\uparrow, n_\downarrow)]}{\partial n_{\sigma}}\right|_{n_\sigma \to n_\sigma(x,t)}~,
\end{equation}
where $\varepsilon^{\rm hom}_{\rm xc}$ is the xc energy (per particle) corresponding to the Yang model ${\hat {\cal H}}_{\rm Y}$, which can be easily found from the Bethe-{\it Ansatz} equations in the thermodynamic limit~\cite{yang_1967,ref:saeed_2007}. 

The {\it dynamical} (or non-adiabatic) contribution to the xc potential, $V^{(\sigma)}_{\rm dyn}[n_\sigma, j_\sigma](x,t)$, is given by
\begin{equation}\label{eq:xc_implicit}
V^{(\sigma)}_{\rm dyn}(x,t)=-\int_{-\infty}^x dx' F^{(\sigma)}_{\rm sd}(x',t)~,
\end{equation}
where $F^{(\sigma)}_{\rm sd}$ is the spin-drag-related force~\cite{scd_giovanni} 
exerted by the atoms with spin $\bar \sigma$ on the atoms with spin $\sigma$,
\begin{equation}\label{eq:force}
F^{(\sigma)}_{\rm sd}(x,t)=-\left.m\frac{n_{\bar \sigma}}{n \tau_{\rm sd}}(v_\sigma-v_{\bar \sigma})\right|_{n_\sigma \to n_\sigma(x,t)}~.
\end{equation}
Due to Galileian invariance this force depends on the relative velocity between the two atom species, $v_\sigma-v_{\bar \sigma}=j_{\sigma}/n_\sigma-j_{\bar \sigma}/n_{\bar \sigma}$.  
In Eq.~(\ref{eq:force})  $\tau_{\rm sd}$  is the spin-drag relaxation time -- the inverse of the rate of momentum transfer between atoms 
of opposite spin orientation -- which has recently been calculated in 1D~\cite{rainis_prb_2008}. 
We will use the results of Ref.~\onlinecite{rainis_prb_2008} as input for our numerical calculations.  
Using Eq.~(\ref{eq:force}) and the continuity equation in Eq.~(\ref{eq:xc_implicit}) we find
\begin{eqnarray}\label{eq:xc_explicit}
V^{(\sigma)}_{\rm dyn}(x,t) &=&-\int^x_{-\infty} dx' \frac{m \left.\tau^{-1}_{\rm sd}\right|_{n_\sigma \to n_\sigma(x',t)}}{n(x',t)}\nonumber\\
&\times &\sum_{\sigma'}\frac{\sigma\sigma' n_\uparrow(x',t)  n_\downarrow(x',t)}{n_\sigma(x',t)n_{\sigma'}(x',t)}{\cal F}_{\sigma'}(x',t)~,
\end{eqnarray}
where ${\cal F}_{\sigma'}(x',t)=\int^{x'}_{-\infty} dx'' \partial_t n_{\sigma'}(x'',t)$.\\

\begin{figure}[t]
\begin{center}
\includegraphics[width=1.0\linewidth]{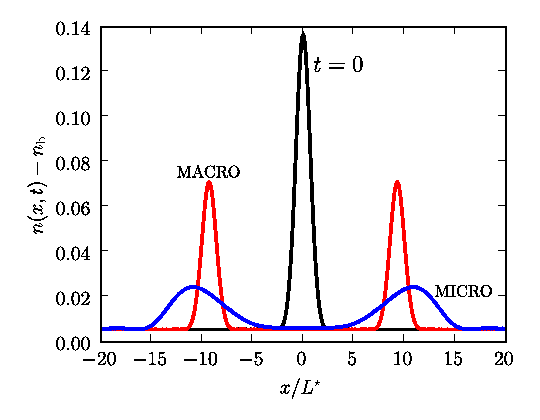} 
\includegraphics[width=1.0\linewidth]{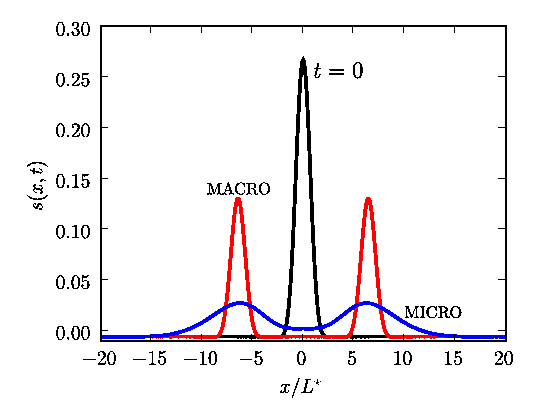}
\caption{(Color online) Top panel: charge density $n(x,t)$ (in units of $1/L^\star$) 
as a function of distance $x$ (in units of $L^\star$), 
after the subtraction of the background constant $n_{\rm b}=N/L$. The initial density profile 
is the wavepacket centered at $x=0$ and labeled by ``$t=0$". The (red) solid line labeled by ``MACRO" represents the corresponding
$n(x,t)$ at a later time $t=5~\hbar/E^\star$, calculated according to the macroscopic theory of Ref.~\onlinecite{polini_2007}. 
The (blue) solid line labeled by ``MICRO" represents $n(x,t)$ at the same time but calculated according to TD-SCDFT 
[Eqs.~(\ref{eq:kss})-(\ref{eq:closure})] {\it without} spin-drag contribution [{\it i.e.} with $V^{(\sigma)}_{\rm dyn}(x,t)=0$]. 
Bottom panel: spin density $s(x,t)$ (in units of $1/L^\star$) as a function of $x/L^\star$. 
Color coding and labeling are the same as in the top panel.\label{fig:one}}
\end{center}
\end{figure}

\begin{figure}[t]
\begin{center}
\includegraphics[width=1.0\linewidth]{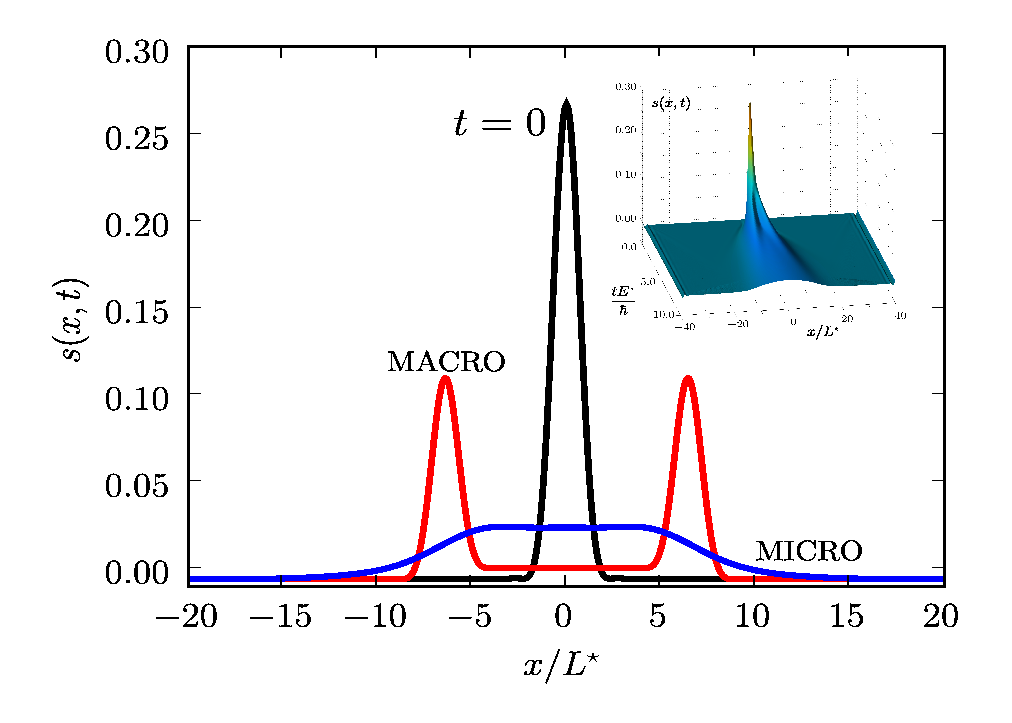}
\end{center}
\caption{(Color online) Same as in the bottom panel of 
Fig.~\ref{fig:one} but {\it with} spin-drag contribution. In the inset we show a 3D plot of the TD-SCDFT result for the 
spin density $s(x,t)$ (in units of $1/L^\star$) as a function of $x/L^\star$ and $t E^\star/\hbar$. \label{fig:two}}
\end{figure}

\noindent {\it Qualitative analysis of spin-charge separation}--- 
Before proceeding with the numerical analysis we want to clarify the mechanism by which the KS equations~(\ref{eq:kss}) 
produce 
independent evolutions of the charge and spin density.  To see this, it is not even necessary to go beyond the ALSDA.  
The essential point is that the KS equation guarantees not only the continuity equation but also the continuity equation for the momentum density, which reads  $\partial_t j_\sigma(x,t) = -m^{-1} \partial_x P_\sigma(x,t)$, where the quantum pressure $P_\sigma (x,t)$ can be expressed in terms of KS orbitals and is therefore an implicit functional of the densities.  
Combining the two conservation laws we arrive at  $\partial_t^2 n_\sigma(x,t)=m^{-1}\partial_x^2P_\sigma(x,t)$, which looks almost like a classical wave equation.  Indeed a classical wave equation is immediately obtained in the LRR (small deviation from homogeneous, unpolarized state) since the quantum pressure can then be approximated, in ALSDA,  as a linear functional of the densities: 
$P_\sigma = \sum_{\sigma'} f_{\sigma \sigma'}\delta n_{\sigma'}$ where $\delta n_{\sigma'}$ are deviations from equilibrium.  Then, after simple algebraic transformation we arrive at two independent wave equations for $n(x,t)$ and $s(x,t)$ 
with two different velocities, $v_{n(s)}= \sqrt{(f_{\uparrow\uparrow}\pm f_{\uparrow\downarrow})/m}$, respectively.  
Admittedly this analysis pertains to the LLR.  Spin and charge are not expected to be truly independent in the nonlinear regime.  But since the correct linear-response limit is built into the KS equation we expect that a strong signature of spin-charge separation will be seen also in the nonlinear regime.  Our numerical calculations confirm this expectation.

\begin{figure}
\begin{center}
\includegraphics[width=1.0\linewidth]{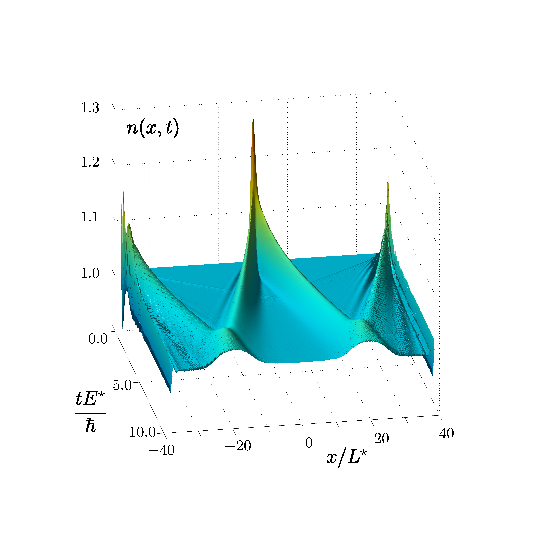}
\end{center}
\caption{(Color online) A 3D plot of the TD-SCDFT result for the 
charge density $n(x,t)$ (in units of $1/L^\star$) as a function of $x/L^\star$ and $t E^\star/\hbar$. \label{fig:three}}
\end{figure}

\noindent {\it Numerical results and discussion}--- 
We have solved Eqs.~(\ref{eq:kss})-(\ref{eq:closure}) with a two-step predictor-corrector Crank-Nicholson scheme. While the scheme outlined above is completely general, for simplicity, in the numerical calculations we have taken a simple box-shaped trapping potential, $V(x)=0$ for $-L/2 < x < L/2$ and $V(x)=+\infty$ elsewhere. We use $L^\star=L/N$ as unit of length 
and $E^\star=\hbar^2/(m L^{\star 2})$ as unit of energy. For definiteness, we fix 
$N_\uparrow=N_\downarrow=40$ ($N=80$ fermions in total), 
$A_\uparrow=-E^\star$, $w=L^\star/\sqrt{2}$, $g_{\rm 1D}=2.0~E^\star L^\star$, and $T=0.5~E^\star/k_{\rm B}$.
In Fig.~\ref{fig:one} we compare the evolution calculated in the ALSDA (which does not include spin drag) with the evolution obtained from the macroscopic theory~\cite{polini_2007}.  
Both theories predict a splitting of the initial peak into two peaks that propagate with different velocities (spin being slower than charge) in agreement with the general picture of spin-charge separation.  However, in the microscopic calculation the magnitude of the peaks decreases far more rapidly with time. It must be appreciated that this happens in spite of the fact that the total spin and particle number are conserved quantities in both approaches.
Note also that the width of the microscopic results in Fig.~\ref{fig:one} is much larger than that of the macroscopic results: the reason is that 
the microscopic theory includes diffusive-like mechanisms related to single-particle thermal excitations [see Eq.~(\ref{eq:closure})], which operate both in the charge and spin channels and are completely missed by the macroscopic theory. Finally, notice the asymmetric forward-leaning shape of the density pulse calculated from the microscopic theory.  This is a nonlinear effect, likely due to the fact that the local velocity, proportional to the density, is higher at the center of the pulse than at its edges~\cite{non_linearity}. 

Fig.~\ref{fig:two} shows the effect of spin drag, 
which enters the macroscopic calculation as a dissipative term in the wave equation for 
$s(x,t)$ [see Eq.~(10) in Ref.~\onlinecite{polini_2007}], the microscopic one through the current-dependent part of the xc potential [Eq.~(\ref{eq:xc_explicit})].  It is evident that the effect of the spin drag is much more pronounced in the microscopic calculation, where the double-peak structure is completely lost.  This happens in spite of the fact that the spin-drag coefficient $\tau_{\rm sd}$ is generally smaller in the microscopic calculation, due to the effect of the spin polarization~\cite{rainis_prb_2008}. Clearly, the inclusion of microscopic excitations weakens the collective behavior of the spin density.  The large difference between the macroscopic and microscopic results demonstrates the importance of relying on the latter when performing quantitative calculations.  

In Fig.~\ref{fig:three} we show a 3D plot of the time evolution of the density packet.  What is notable here is that already at short times some density waves appear, coming from the sharp edges of the waveguide: at larger time they mix with the original packet. These Friedel-like oscillations are completely beyond the power of the macroscopic approach. In conclusion, the above calculations amply demonstrate the versatility of the TD-SCDFT method in producing detailed results which can be compared with future experiments on the propagation of charge and spin pulses in ultracold Fermi gases.

\noindent {\it Acknowledgments}--- G.X. was supported by NSF of China under Grant No. 10704066 and by DOE Grant No. DE-FG02-05ER46203. 
G.V. was supported by NSF Grant No. DMR-0705460.


\begin{thebibliography}{77}
\bibitem{accum}
	We have assumed a simple Gaussian form for $W_\sigma(x)$ only for definiteness: the TD-SCDFT 
	scheme outlined below is completely general.
\bibitem{zeroT}
	In writing Eq.~(\ref{eq:alsda}) we are neglecting the temperature dependence of 
	the ALSDA xc potential, which we expect to 
	be unimportant for the physics we are describing here.
\bibitem{non_linearity}
	This effect is absent in the spin channel, confirming that spin-charge separation holds with good accuracy in spite of nonlinearity.
\end{thebibliography}
\end{document}